\documentstyle[eqsecnum,preprint,aps,amssymb]{revtex}
\begin{document}
\title {Dynamic transitions  and hysteresis} 

\author {Bikas K. Chakrabarti$^{a,*}$ and Muktish Acharyya$^{b,c,+}$}

\address {$^a$Saha Institute of Nuclear Physics, 1/AF Bidhannagar, Calcutta-700 064, India \\
$^b$Institute for Theoretical Physics,  University of Cologne, D-50923 Cologne, Germany}

\date{\today}
\maketitle

\begin{abstract}

\noindent When an interacting many-body system, such as a magnet, is
driven in time by an external perturbation, such as a magnetic field,  
the system cannot respond instantaneously due to 
relaxational delay. The response of such a 
system under a time-dependent
field leads to many novel physical phenomena with intriguing 
physics and important technological applications. For oscillating
fields, one obtains  hysteresis  that would not occur
under quasistatic conditions in the presence
of thermal fluctuations. Under some extreme conditions
of the driving field, one can also obtain a non-zero average
value of the variable undergoing such ``dynamic hysteresis''. This
non-zero value  indicates a breaking
of symmetry of the hysteresis loop, around the origin. Such a 
transition to the ``spontaneously broken symmetric
phase'' occurs dynamically when the 
driving frequency of the field increases beyond its threshold
value which  depends on the field amplitude and the temperature. Similar
dynamic transitions also occur for pulsed and stochastically varying 
fields. We present an overview of the ongoing researches in this
not-so-old field of dynamic hysteresis and transitions.
\end{abstract}


\bigskip

\leftline {\bf  Contents}

\leftline {I. Introduction}
\leftline {II. Magnetic response under an oscillating field:} 
 
Model studies

\leftline {III. Dynamic hysteresis}
\leftline {IV. Dynamic transitions}

A. Due to oscillating fields

B. Due to pulsed and stochastic fields

(a) Response to a positive pulse field

(b) Response to a negative pulse field and the 

magnetisation-reversal transition

(c) Dynamic transition due to random 

(time varying) fields

\leftline {V. Concluding remarks}

\bigskip

\section {Introduction}

\medskip

\noindent Consider a co-operatively interacting many-body system, such 
as a magnet, driven by an oscillating external perturbation, such as an 
oscillating magnetic field. The thermodynamic response of the system,
e.g., the magnetisation, will then also oscillate 
with necessary modifications
in its form, and will lag behind the 
applied field due to the relaxational delay.
This delay in the dynamic response gives rise to a nonvanishing 
area of the magnetisation-field loop, which phenomenon we term
{\it dynamic hysteresis}.
When the time period of oscillation of the 
external perturbation becomes much less compared to
the typical relaxation time of the thermodynamic system,
the hysteresis loop becomes asymmetric around the origin and
an interesting thermodynamic phase arises 
{\it spontaneously} out of {\it 
dynamically broken symmetries}
due to the competing time scales in such nonequilibrium driven systems.

In the example of a magnetic system, the time ($t$) variation of
the magnetisation $m(t)$ lags behind that of the oscillating field $h(t)$
(= $h_0 \sin \omega t$, say), and after some initial transient
period, the dynamic   
$m(t) - h(t)$ loop stabilises and encloses a nonvanishing
loop area $A(T, h_0, \omega)$ [= $\oint m~dh$], which
 depends on the temperature
$T$ of the system and the field amplitude $h_0$ and frequency $\omega$.
This hysteresis is dynamical
in origin and disappears in the quasistatic limit.
Pure magnetic systems, without any random defects or anisotropies 
to pin the magnetic domains, can relax properly in the 
quasistatic limit and follow the field in phase due to 
the presence of thermal fluctuations at any finite temperature.
No hysterisis can  therefore occur in pure magnets in the 
quasistatic limit (Landau and
Lifshitz 1935, Feynman
et al 1964, Kittel 1966). It may be mentioned here that a 
subset of the
engineering studies on hysteresis are of course in this quasistatic limit, in 
materials containing random pinning defects (see e.g., the Proc. Workshop
on Hysteresis Modelling and Micromagnetism, Physica B {\bf 233} (1997) 259).
Such quasistatic hysteresis in random materials has been modelled
recently by using magnetic models having randomly quenched magnetic fields
(Mirollo and Strogatz 1990, Sethna et al 1993).
One studies here the self-organising avalanches in the zero-temperature
spin-flip dynamics under external field (see e.g., Sethna et al 1993,
Dahmen and Sethna 1993, Dhar et al 1997). 
For most of the design engineering problems  connected with the 
recording processes and materials, the study of hysteresis is essentially
dynamic in nature (see  Torre 1966, Mallinson, 1987). A power law dependence
of such hysteretic loss  
on the  magnetic induction (related to the external field)
 was first proposed
empirically,  more than a century ago, by Steinmetz (1892).
Here, in this colloquium, we
confine ourselves to  pure magnetic systems and examine the dynamic
hysteresis
arising out of two competing time scales, namely the 
relaxation time of the system
and the time period of the external field. As such, it exists
in any pure extensive (cooperative) system
at any finite temperature, and disappears in the static limit. Following
an early study of the dynamics
in a driven model bistable
system having rate competition (Agarwal and Shenoy 1981), the low 
frequency and amplitude power law behaviour for such dynamic hysteretic
loss was first investigated in a multicomponent magnetic
model by Rao et al (1989, 1990a), in the limit of a large number of
components.

An important aspect of the magnetisation-field loop or
$m-h$ loop shape, in this
example of the dynamics of a pure magnet, is 
that the loop becomes asymmetric (in the positive and negative ranges of 
$m$) as the
driving frequency of the field increases. The reason 
for this  is that the
system
does not get enough time to relax, even to follow the
sign (phase) of the external field. This causes 
a spontaneous symmetry breaking
and a nonvanishing value of the dynamic order parameter $Q$ given
by the time averaged magnetisation over a complete period ($Q =
{\omega \over 2 \pi} \oint m ~dt$). 
This non-zero value of the order parameter $Q$
develops spontaneously in the high frequency ranges, although the
external perturbation does not provide, on average (over time), any
symmetry breaking field ($\oint h~dt$ = 0). A prototype of this
dynamic transition was  first observed  by Tom$\acute{\rm e}$
 and de Oliveira (1990) 
in the numerical solution of
the mean field equation of motion   
of the classical one component magnetic model
(see later in the next section). However, the transition
there is not truely dynamic as it can exist 
for such  equations of motion even in the 
zero frequency (static) limit of the driving field. This transition in the
static limit is an artifact of the mean field approximation, which
neglects nontrivial fluctuations. The 
occurrence of the dynamic transition 
for models incorporating thermodynamic
fluctuations, was later shown in several Monte Carlo simulation
studies (Lo and Pelcovits 1990, Acharyya and Chakrabarti 1994, 1995). 
In these numerical studies, incorporating fluctuations, the
dynamic transition disappears in the static limit.

Although the subject of nonequilibrium phenomena associated with 
first order phase transitions has received considerable attention in
the recent past, most of the attention has been focussed on 
two variants of the problem:  Rapid
quenching of a melt or of a spin system from a disordered phase to one
of the competing ordered phases
(see e.g., Bray 1994), or to the century-old problem
of the decay (often remarkably slow, as for example in some allotropic
forms of carbon) of the  metastable phases of a condensed matter
system through the nucleation of the domains of various phases, as the
external conditions of the system are suddenly changed (see e.g., 
Rikvold and Gorman 1994).  The interest in the study
of nonequilibrium phenomena associated
with extended  systems, driven periodically or stochastically
between two (or more) equivalent
ordered phases, is somewhat more recent. These phenomena are
also related to the phenomena of {\it stochastic resonance} in bistable
systems which are periodically driven in the presence of stochastic 
noise (see e.g., Gammaitoni et al 1998). Here, the resonance of the weak
periodic modulation (which by itself is incapable of inducing any
phase change)  with the Kramers' frequency of the theromodynamic
system (coming from the finite temperature
 Boltzmann probability of activated hopping across
a finite barrier between the two ordered phases) finally succeeds
in inducing the periodic phase changes of the entire macroscopic
system. Since its early introduction as a possible model for the
dependence of the earth's ice age periods on the weak periodic 
modulations of the earth's dynamics through resonance of the
stochastic noise of its weather (Benzi et al 1981),
the concept has found wide applications to a variety of physical
phenomena in biological and engineering sciences, including to
dynamic hysteresis (Sides et al 1998a).

Interestingly enough, in spite of its immense practical importance
in the engineering problems of magnetic recording, and the prospect
of very interesting physics, the problem 
of dynamic transition and hysteresis
in periodically driven magnetic systems was overlooked for many
years. Although it is now less than ten years' old, the subject
has become a field of considerable novelty and vitality. A number of
key issues are already settled and a few others are nearly resolved. 
In this  Colloquium we try to give an introduction 
to these intriguing nonequilibrium dynamic transition phenomena in 
driven extended systems and to discuss simple analytic theories
connecting them. 

\bigskip

\section {Magnetic response under an oscillating
 field:
Model studies}

\medskip

\noindent The detailed nature of the dynamic response of extended systems 
(having many interacting degrees of freedom), under time-dependent 
fields, is  being investigated intensively these days. 
Considerable efforts have been
 made, in particular using computer simulations,
 to investigate the nature of the
above mentioned dynamic phase transition and hysteresis in Ising models.
Simple Ising systems contain ferromagnetically 
interacting spin degrees of freedom, each
with binary (up/down or $\pm 1$) spin states.
Let us consider a simple ferromagnetic
system  represented here by an 
Ising model with nearest neighbour ferromagnetic coupling, which is put
under an oscillating external field. Such a system
can be represented by the Hamiltonian

$$H = -\sum_{<ij>} J_{ij} s_i s_j - h(t) \sum_i s_i, \eqno (1)$$

\noindent with

$$ h(t) = h_0 \sin \omega t. \eqno (1a)$$

\noindent Here $s_i ~(=\pm 1)$ represents the  Ising spin variable at
the site $i$ on a $d-$ dimensional lattice 
and $J_{ij}$ represents the spin-spin interaction
strength  between sites $i$ and $j$. $\sum_i$ runs over all the
lattice sites and $\sum_{<ij>}$ runs over all the 
distinct nearest-neighbour
pairs. Note that for truly long-range interactions, as in the mean
field models discussed later, the sum in (1) extends over
all the pairs and the value of $J_{ij}$ decreases inversely with the
number of sites in the system. 
The system is in contact
with an isothermal heat bath at temperature $T$. For simplicity
all $J_{ij} (> 0)$ are taken equal to  a constant $J (= 1)$, 
and the temperature $T$ is
measured in units of $J$, setting the  
Boltzmann constant at unity.
In order to keep $H/T$ dimensionless, $h_0$
in (1a) is also measured in units of $J$.

We will discuss later the equations of 
motion for the average magnetisation
$m (= \langle s_i \rangle,$ where $\langle ...\rangle$ denotes the
ensemble average) for such systems. It may be noted here that
since no transverse magnetisation is possible for the Ising system (the 
magnetisation $m$ is a scalar), the magnetisation can only try to
follow the field, albeit with a delay. The behaviours of the
 response magnetisation $m(t)$
 are qualitatively indicated in Fig. 1 for some typical cases.
Generally, since the equation of motion of 
such systems remains invariant
for $t \rightarrow t + 2\pi/\omega$, 
the response magnetisation also becomes
periodic (not necessarily sinusoidal, even though the field is so) with 
the same periodicity or some integer multiple of it (in cases
of stochastic resonance). This is responsible for the feature that  the
magnetisation $m$ can at most be double-valued at any field $h$,
 and the $m-h$
loop has to be closed
(on average; in the thermodynamic limit).
 The phase delay in $m(t)$ (compared to that of
$h(t)$) gives rise to the dynamic hysteresis loss or the loop area $A $.
The same delay and the constraint of identical periodicity for the field
and the response magnetisation 
can induce an  asymmetry in the response (in
$\pm m$) as the driving frequency increases. 
When  the time period of oscillation
falls far behind the effective relaxation time of the many-body system,
a dynamically broken symmetric  phase
 arises spontaneously with a nonvanishing value of the
dynamic order parameter $Q$ (the period averaged magnetisation). 

Typically, for fixed temperature $T$ and field amplitude $h_0$, 
the dynamic hysteresis loss $A$ increases with increasing frequency
$\omega$ for low values of 
$\omega$. This is because, for low values of $\omega$,
the effective delay in the response increases as $\omega$ increases. In
general, for a fixed $\omega$,  $A$ increases with decreasing $T$ 
and increasing $h_0$  until $A$ saturates.   
Eventually, as the driving frequency exceeds a threshold value (dependent
on $h_0$ and $T$; see Acharyya 1998b), the loop area $A$ starts decreasing, 
because of the increase in the effective delay (phase lag)
towards $2\pi$. Eventually, the loop
area vanishes for very high frequencies when the  dynamic symmetry is 
fully broken ($Q = 1$). For
a fixed frequency, this dynamic transition {\it
phase boundary} (seperating
non-zero and zero values of $Q$
 in the $h_0 - T$ plane) is in general convex
towards the origin. With large values of the 
field amplitude $h_0$ or temperature $T$, one gets a ``forced oscillation''
kind of scenario inducing the dynamically symmetric ($Q =0$)
phase. However,  right near the dynamic phase boundary, in the symmetric
($Q = 0$) phase, the field amplitude $h_0$ may not be sufficient to
cross the free energy barrier at that
temperature and stochastic resonances 
can induce such dynamic symmetry 
in several attempts or in averages over several cycles (Sides et al 1997,
1998a).
With decreasing $\omega$, the
phase boundary shrinks inward. Finally, in the static limit ($\omega =0$)
the dynamic transition disappears and the phase boundary in the $h_0-T$
plane collapses to a line with $h_0 = 0 $ and ending at $T = T_c$, the
static transition temperature of the unperturbed system. 

The majority of  studies on such Ising systems in an oscillating field have
been made employing the Monte Carlo method (see e.g., Binder 1979) using
the Metropolis single spin flip dynamics. 
Starting from an arbitrary initial
state or configuration of spins (e.g., with 
all spins up), the spin state $s_i(t)$
at each site $i$ and time $t$ is updated 
sequentially with a probability  proportional to
exp[$-{1 \over T}\Delta E_i
(t)]$ where $\Delta E_i(t) = 2 s_i[\sum_j s_j(t) + h(t)]$, the change
in energy due to the spin flip.
One full scan over the entire lattice is defined as 
the unit time step (or Monte
Carlo step per site). The response magnetisation (per site) at time $t$
is then easily calculated: $m(t) = (1/N) \sum_{i} s_i(t)$,  where $N$ is 
the total number of spins in the lattice ($N = L^d$ if one considers a
$d$-dimensional hypercubic lattice of 
linear size $L$). One then concentrates
on the behaviour of the response function $m(t)$ for a long time $t$ (much
after the stabilisation, starting from the 
initial state) as compared to the
field $h(t)$. Specifically, one considers 
quantities like the loop area $A (=
\oint mdh)$ and the dynamic order 
parameter $Q (= {\omega \over 2\pi} \oint
mdt)$. 

The  mean field{\footnote [1] {In a 
cooperatively interacting system as, for example, described by the
Hamiltonian (1), each degree of freedom (here spin) interacts with
the neighbouring one. Hence the interaction tries to induce complete
order ($m$ = 1) in the system. The thermal noise, at any nonvanishing $T$,
induces fluctuations destroying the  order and takes the system to
microscopic states which are not energetically favourable. 
The probabilities
of such higher energy states are given by the Boltzmann probabilities
 at
that $T$. The overall order is then given by the thermodynamic
average value ($m < 1$), while there are fluctuations around this average.
In the mean field approximation, one isolates an arbitrarily chosen
cluster of spins (usually one spin) and performs the statistical 
mechanics for these chosen spins exactly, taking the states of
all its (interacting) neighbours at their mean or average state (represented
by the average $m$). This self-consistent
approximation therefore pictures each
degree of freedom to be placed in a mean field provided by the average
state of the neighbours and the interaction strength $J$ (in (2)
the value of $T_c = JZ_{nn}$
has been put equal to unity, where $Z_{nn}$ denotes the number of nearest
neighbours of any spin).}}
 equation of motion for the 
average magnetisation $m$ can be written as

$$ {dm \over dt} = -m + \tanh \left({m + h(t) \over T}\right), \eqno (2)$$

\noindent where $h(t)$ is given by (1a) and the microscopic relaxation
time (on the right hand side of (2)) is put equal to unity (see
Suzuki and Kubo 1968). This simple nonlinear equation
(in one variable) is indeed
capable of capturing a number of important features of
dynamic hysteresis  and of the dynamic
transition. However,
the  lack of fluctuations 
of the thermodynamic average value $m$ in the above
equation is responsible for  the loss of some
very significant features. For example, even in the quasistatic limit,
one requires a nonvanishing coercive field amplitude to overcome the
free energy barrier or  to go from a $+m$ 
state to the corresponding 
$-m$ state crossing
the barrier which is finite below the static order-disorder transition
temperature $T_c$. 
Consequently, a nonzero hysteresis loop area can be found even in the
static limit.
Also, any applied
field here (of magnitude below that of the coercive
field which is nonzero for $T < T_c$) cannot break the 
asymmetry of the
system.
The ``dynamic transition'' therefore occurs here even in the static limit! 
Still, considerable analytical studies 
have been made for hysteresis from such
single-variable, nonlinear, periodically-driven 
equations of motion or maps
(Jung et al 1990, Goldsztein et
al 1997). Some efforts have  been made 
to incorporate
the effects of Gaussian  noise 
on the Langevin type equation of motion for $m$ with 
a somewhat weaker nonlinearity (Mahato and Shenoy 1994, Paniconi
and Oono 1997).
 Efforts have also been made to treat the effects of
fluctuations using a renormalisation group technique for a general
$n$-component system (with $n$ components for the magnetisation vector
$m$), and study the  $n \rightarrow \infty$ limit (Rao et al 1990a,b,
Rao and Pandit 1991, Thomas and Dhar 1993, Fan and Jinxiu 1995).

\bigskip

\section { Dynamic hysteresis}

\medskip

\noindent The dynamic contribution to the coercive field and the
hysteresis loop area has been investigated in several recent
experiments, mostly in thin films or in two dimensions.
 Bruno et al (1990) studied the dependence of the hysteresis
loop area $A$ on the rate of change of the external field,
 in ultrathin ferromagnetic films.
Their study gives some indirect 
information on the dynamic contribution to the loop area $A$.
 In a more recent experiment, Jiang et al (1995, 1996a,b) studied the
frequency-dependent hysteresis of epitaxially 
grown ultrathin (2 to 6 monolayer
thick) Co films on a Cu(001) surface at room temperature. 
The films have
strong uniaxial magnetisation with two ferromagnetic phases of
opposite spin orientations. This  magnetic 
anisotropy  makes it appropriate to represent the system
by an Ising-like model
(see also He and Wang 1993). The external
magnetic field $h(t)$ on the system was driven sinusoidally in the 
frequency ($f = \omega /2\pi$) range 0.1 to 500 Hz and in the amplitude
($h_0$) range 1 to 180 Oe. Here of course the time-varying current or
the magnetic field induces an eddy current in the core, 
which results in a 
counter-field reducing the effective magnitude of the applied field. 
The surface magneto-optical Kerr effect technique 
was used to measure the
response magnetisation $m(t)$. A typical variation of the 
loop area $A$ with the driving frequency $f$, at  room temperature
and at fixed external field amplitude $h_0$, is shown in Fig. 2a. 
Also, it may be mentioned that in a recent similar experiment on dynamic
hysteresis in ultrathin Fe films on W(110) surface (Suen and Erskine
1997), the typical behaviour of the dynamic hysteresis
is observed to be similar,
 although considerable discrepancies 
are observed in the actual details
for different materials and regimes.

The observed variation of the loop area $A$ with frequency $\omega$
follow the generic form discussed earlier: $A$ decreases for both low  and 
high values of $\omega$. However, it may be noted that $A$ does not quite
vanish in the zero frequency limit. The observed variation can, in fact, 
be fitted to a form

$$A = A_0 + h_0^{\alpha} \omega ^{\beta} g\left({\omega \over h_0^{\gamma}}
\right), \eqno (3)$$

\noindent with the scaling exponents $\alpha, \beta$ and   $\gamma$
and with the scaling
function $g$ having  a suitable
nonmonotonic form such that $g(x) \rightarrow 0$ as $x \rightarrow 0 ~{\rm
or}~ \infty$.{\footnote [2] { The constants $\alpha,
 \beta, \gamma$, etc, in general
assume noninteger values, indicating singularities in the power laws. 
Very small values of the constants are often taken as indications
for logarithmic variations rather than power laws (taking the 
representation ln$x$ = [($x^n -1)/n]_{n \rightarrow 0}$). These
constants are called the scaling exponents here, in conformity with the
practice in critical phenomena, where the exponents 
also turn out to be universal
in the sense that they do not depend on 
most of the details of the thermodynamic
system. The scaling funcition is again a term borrowed from the
critical phenomena literature, where the function $g$ is not dependent
on the individual values of $\omega $ and $h_0$, but is a function 
only of the scaled variable $\omega / h_0^{\gamma}$.}} 
Here $A_0$ is the loop area in the zero frequency limit.
 It seems, depending on the nature of the dynamic
processes involved in different materials and also the ranges (of 
amplitude, frequency, etc), the values of the exponents differ
dramatically. While Jiang et al (1995) obtained $\alpha \simeq 0.67 \simeq 
\beta $ for Co films on Cu(100), and $\alpha \simeq
0.6 $ and $\beta \simeq 0.3$ for Fe films on Au(001), Suen and Erskine (1997)
obtained much lower values for the same exponents: $\alpha \simeq 0.3$
and $\beta \simeq 0.06$ (perhaps logarithmic) for Fe 
films on W(110) surfaces.
 It may be mentioned that
Jung et al (1990) observed, in the context of
their analytical study for a model switching between two modes
of a semiconductor laser,  that one needs to replace $h_0$ in (3) by
$\tilde h_0$, to  account for a subtraction of a threshold field
amplitude below which no hysteresis takes place. 

This experimental observation perhaps indicates that the 
observed hysteresis is not entirely dynamic in origin; $A$ does
not quite vanish in the static limit $\omega \rightarrow 0$. An
approximate mean field solution of (2)
can of course  give the above fitting form (3) for $A$ (Jung et al 1990,
Luse and Zangwill 1994, Hohl et al 1995, Goldsztein et al 1997):
Linearizing the mean field equation (2), for small $m$ and $h_0$ ($T > 
T_c = 1$), one gets

$$ {dm \over dt} = -\epsilon m + { h(t) \over T}; ~~~ \epsilon = 
{T - 1 \over T}. \eqno (4) $$

\noindent The steady state solution of which 
 can be written as $m(t) = m_0 \cos (
\omega t - \phi)$ for $h(t) = h_0 \cos \omega t$. A direct substitution
then gives
$ m_0 = {h_0 /T \sqrt{\epsilon^2 + \omega^2}} ~~~ {\rm and} ~~~
\phi = \sin^{-1} ({\omega / \sqrt{\epsilon^2 + \omega^2}} ).$
For the loop area $A$ in this linearised limit, one gets

$$ A = \oint m dh \sim 
h_0^2 g(\omega)/T;$$ 
$$g(\omega) = {{ \omega } \over  {\epsilon^2 + \omega^2}}. 
\eqno (5) $$

\noindent The above Lorentzian form for the variation of the loop area
$A$ with  frequency $\omega$  is in fact valid for $T > T_c$. For
$T \lesssim T_c$ a similar form for the frequency variation will be valid
with an additional  frequency-independent contribution $A_0(T) \sim
m_e(T)h_c(T) \sim (T_c - T)^2$, where $m_e(T)$ denotes the equilibrium
magnetisation and $h_c(T)$ the static coercive field at that temperature $T
< T_c ~ (= 1$ here). The above Lorentzian form (5) for the variation 
of the dynamic hysteresis loop area $A$ with the frequency $\omega$ of
the driving field, at high temperatures, may be compared with that given
in (3) with $\alpha = 2, \beta = 1$ and $\gamma = 0$. In fact, the
numerical solution for the loop area $A$  from (2) for $T > T_c ~(= 1)$
gives excellent agreement with (5) ($A_0$ is zero for $ T > T_c$). 
At higher field amplitudes and lower temperatures
($T < T_c)$, the above linearisation of the  tanh term in (2) becomes
inappropriate. Jung et al (1990) and Goldsztein et al (1997)
 considered the effect of an additional $m^3$ 
term  in the right-hand side of (4). In the limit $\omega \rightarrow 0$, 
they could solve such a nonlinear equation and obtained $A \sim A_0 +
\tilde h_0^a \omega^b$ with $a = 2/3 = b$ for positive
values of  $\tilde h_0 = h_0 - h_c $.

 In   models for order-disorder ferroelectrics,
the co-operative dipole-dipole 
interaction is represented by an Ising Hamiltonian, while the quantum
tunnelling between the two equivalent wells (corresponding to Ising states
$\pm 1$) is represented by a (non-commuting) tunnelling field.
A similar mean field treatment of the semi-classical equation of motion of
a quantum (Ising) system, for finite temperature hysteresis, was performed
by Acharyya et al (1994). Here  the tunnelling 
field is driven sinusoidally to simulate a  periodic modulation of
the external pressure on the sample.

All these mean field  
studies, mentioned above, are for systems without any thermal
fluctuations. In any realistic system, as in the experimental systems
described above, such fluctuations are present and theoretical
analysis becomes considerably 
more difficult. However, computer simulation 
studies on such models are possible and these studies give significant
insights regarding the effects of such fluctuations. 
Extensive Monte Carlo studies have been made for the Ising system
represented by the Hamiltonian (1) in one to four dimensions (Lo and
Pelcovits 1990, Acharyya and Chakrabarti 1994, 1995).
Acharyya and Chakrabarti observed that at high
temperatures ($T > T_c$) the loop area variation for the entire
frequency range can be represented by

$$A(T, h_0, \omega) \sim h_0^{\alpha} T^{-\rho} g \left({\omega
\over{h_0^{\gamma}T^{\delta}}}\right), \eqno (6)$$

$$ g(\tilde \omega) \sim \tilde \omega^{\beta} \exp 
(-\tilde \omega^2/\sigma), \eqno (6a)$$

\noindent  where the scaling function $g(\tilde 
\omega)$ is exponentially decaying in such 
Monte Carlo studies, with an initial
power-law growth (compared to the Lorentzian form 
(5) in the linearised mean
field case). The above scaling form (6) obviously reduces to a power law
in the low frequency limit: 
$A \sim h_0^a \omega^b T^{-c}$, with $a = \alpha -
\beta \gamma$, $b =  \beta$ and
$c = \rho + \beta \delta$. The fitting curve for
the collapsed  data for the loop
area $A$ as function of the scaled frequency $\tilde
\omega (= \omega /h_0^{\gamma} T^{\delta})$ gives the scaling function
$g(\tilde \omega)$. 
The best fit values of the exponents $\alpha, \beta, \gamma$ and
$\delta$ depend on the dimension. One typical  result for the data
collapse for $A$ in two dimensional Monte Carlo study is shown in Fig. 3.
The best fit values for these exponents  seem to be $\alpha \simeq 1.0$;
$\beta \simeq 0.3$, 0.5 and 0.5; 
$\gamma \simeq 0.9, ~ 0.7 $ and 1.4;
$\delta \simeq 1.2, ~1.8$ and 1.4; and 
$ \rho \simeq 0.8,~ 1.2$ and 0.5, 
in $d $ = 2, 3 and 4 respectively. 
This gives  $a \simeq$ 0.7, 0.6 and 0.3, and
$b \simeq$ 0.3, 0.5 and 0.5 respectively in the above dimensions (Acharyya
and Chakrabarti 1995). For linearly swept fields, extensive simulation 
studies  to check the above scaling behaviour of $A$ has been done
recently by Zheng and Zhang (1998a,b). They estimated 
the value of the dynamic exponent $z$ (see section IV B), from the
exponent $\beta$ and compared that with the direct Monte Carlo
estimates for the same.

If the applied field amplitude and the system size are both small, 
one can utilise a picture for the spin flips occurring through the
{\it nucleation} of a single spin domain (Thomas and Dhar 1993, Sides et
al 1998a). The classical nucleation theory of Becker and D\"oring (see
e.g., Rikvold and Gorman, 1994) suggests that the nucleation rate is
$I \sim  \exp [-F(l_c)/T]$, given by the optimality
condition $l_c = [\sigma(d - 1)/2d]/h$ of the free energy $F(l) =
2hl^d + \sigma l^{d-1}$ for the formation of a droplet or 
domain of linear
size $l$ under field $h$. Here  $\sigma$ is 
proportional to the surface tension for the formation of the
droplet. For low temperatures the magnetisation 
switches from about +1 to $-1$ as the field is swept from positive
to negative values. The hysteresis loop area $A$ is then 
essentially proportional to the (dynamic) coercive field $h_c \simeq
h_0 \omega t_s$, where the
 switching time $t_s$ can be estimated by requiring  
an order of unity value for the total integrated switching probability
within $t_s$:

$$\int_0^{t_s} dt \exp \left(-{1 \over [h(t)]^{d-1}}
\right) $$
$$ \simeq
(h_0 \omega)^{-1} \int_0^{h_c} dh \exp \left(- {1 \over h^{d-1}}
\right)  = 1.$$

\noindent This suggests that

$$ A \sim h_c  \sim \left[-{\rm ln}
(h_0\omega) \right]^{-1/(d-1)},
\eqno (7)$$

\noindent for small $t_s$ and $\omega$. In this estimate,  the
loop area disappears logarithmically in the static limit.
 Extensive simulation studies by Sides et al (1998a,b) and by  
 Acharyya and Stauffer (1998) 
 for small system sizes
and low field amplitudes seem to suggest that the above 
logarithmic dependence on frequency can indeed be observed for
extremely low frequencies. However the inaccessibility of 
such really low frequencies, both in simulations and in experiments,
and power law type fits to the data for rather high frequencies might
be responsible for the apparent indication of an effective constant value
$A_0$ for the ``extrapolated'' loop area in the zero frequency limit, as
discussed earlier in (3). Let us remember that
 such an analysis is valid only  when a single
domain of flipped spins grows and 
induces the switching in magnetisation.
As the field amplitude increases, and/or the system size becomes much 
larger than the single critical domain size $l_c$, the switching
occurs by the coalescence of multiple  domains. The extension  of
the above nucleation rate analysis from a  single domain to the
crossover region of multi-domains has also been done recently
(Sides et al 1998b,d; see also Sides 1998).
In such strong field 
cases, however,  the power law scaling behaviour
(6) seems to be quite appropriate as indicated by the results
from various Monte Carlo studies (Acharyya and Chakrabarti 
1995).  Analytically, the
effect of such fluctuations can still be tackled in the
large $n$ limit of an $n$-vector
model (Rao et al 1990a), for which one finds $A \sim h_0^a \omega^b$ for
$\omega \rightarrow 0$ in the high temperature limit, with $a = 1/2 = b$
with logarithmic corrections (Dhar and Thomas 1992, see also Fan and
Jinxiu 1995).

\section {Dynamic transitions}

\subsection {Due to oscillating fields}

\noindent As mentioned  in the preceding section,  the experiment by 
Jiang et al (1995) on dynamic hysteresis in
epitaxially grown ultrathin Co films on a Cu(001) surface
with magnetic anisotropy at room temperature 
exhibits a prominent signature of the
dynamic transition  as the driving frequency increases. 
The observed  hysteresis loops tend
to become asymmetric about the zero magnetisation line (see Fig. 2a)
in the high frequency regime. 
Even for a very low frequency ($ f = 4$ Hz), they observed the 
same symmetry breaking
dynamic transition by reducing the amplitude of the magnetic field (see
Fig. 2b).
However, the precise experimental phase boundary for this
dynamic  transition is not available to date. Moreover, no
experimental attempt has been made so far to probe
the thermodynamic nature of such dynamic transitions. Similar a dynamic 
transition can also be clearly seen from the data obtained
by Suen and Erskine
(1997) in Fe films on W(110) surfaces.

\bigskip

\noindent {\bf Mean field scenario:} 

\noindent Tom$\acute{\rm e}$ and de Oliveira (1990) studied the
response of a kinetic Ising model in 
the presence of a sinusoidally oscillating
magnetic field, by solving the  mean field equation of motion (2)
for the average magnetisation. 
The dynamic order parameter $Q $
(the time average magnetisation over a full cycle of the oscillating field),
which vanishes for the symmetric hysteresis loop, was found to assume
nonzero values for some range of values of 
field amplitude ($h_0$) and temperature ($T$), dependent on the 
frequency ($\omega$).
The $h_0-T$ plane 
is then divided by a phase boundary line which separates
the dynamically disordered ($Q = 0$) phase from a dynamically ordered
($Q \neq 0$) phase
for any fixed 
frequency $\omega$. Tom$\acute{\rm e}$
 and de Oliveira (1990) also identified a tricritical
point on the phase boundary line which separates the continuous/
discontinuous nature of the transition
along the phase boundary. A schematic diagram 
of the dynamic phase boundary is shown in Fig. 4.

However, as mentioned before, in this mean field approximation
the dynamic transition can exist 
even in the static limit!  The reason is that for  field amplitudes
less than the 
static coercive field $h_c$ (which is nonzero below the order-disorder
transition temperature $T_c$), the response magnetisation varies
periodically but asymmetrically 
even in the zero frequency limit. The system
then remains locked to the higher, yet locally attractive, 
well of the free energy and cannot go to the other (deeper) well,
unless driven by any noise or fluctuations which are absent in the 
mean field system. 

This mean field dynamic transition phase boundary in the static limit
can be easily estimated. This is because, the $h_0(T)$ line in the 
$h_0-T$ plane for $\omega = 0$ corresponds to the temperature 
dependence of the static coercive field $h_c(T)$. Since the Landau (mean
field) free energy grows as $(T_c - T)^2$ for $T < T_c$, and the
spontaneous magnetisation 
is $m_0 \simeq (T_c- T)^{1/2}$, the coercive
field can be estimated from the balance of $m_0h_c$ with the free energy
barrier height: $m_0 h_c \sim (T_c - T)^2$, or $h_c \simeq
(T_c - T)^{3/2}$. The mean field dynamic phase transition boundary 
(Acharyya and Chakrabarti 1994) indeed
 converges to such a behaviour in the static limit: $h_0(T) \simeq
(T_c - T)^{3/2}$.

\bigskip

\noindent {\bf Thermodynamic nature of the transition:}

\noindent Lo
and Pelcovits (1990) first attempted to study the dynamic
transition in the presence of  fluctuations 
as in the kinetic Ising model, using Monte Carlo simulations. While
they could detect the transition, 
they could not obtain any precise phase boundary
for the transition. Acharyya and Chakrabarti
(1994) obtained later  the dynamic transition 
 phase boundary.
 Afterwards, various studies have been made to investigate the thermodynamic
nature of this transition.

Extensive Monte Carlo studies in two and three dimensions show that a
precise phase boundary  (in the $h_0 - T$ plane)
exists for the transition at any fixed nonzero frequency ($\omega$)
of the driving field. For ($h_0, T)$ values below this boundary, one
gets asymmetric $m-h$ loops (or asymmetric dynamic hysteresis loops), 
while for values above this boundary the loops are symmetric. This
spontaneous symmetry breaking transition of the (dynamic) hysteresis
loop can be accurately described by the behaviour of the 
dynamic order parameter $Q$, which measures the stable (long time)
average value of the magnetisation over a complete period. 
As mentioned before, 
this dynamic breaking of symmetry arises due to the competing time scales
of the oscillating field and that of the response magnetisation.  With
decreasing frequency, the phase boundary line in the $h_0-T$ plane
shrinks towards the origin, and eventually it becomes a line along
the $T$-axis ($h_0 = 0$), ending at $T = T_c$. Recently, Acharyya
(1998b)
identified this dynamic transition point as the point where the
correlation between $h(t)$ and $m(t)$ goes to a minumum.

The Monte Carlo studies also indicate that the transition clearly 
becomes discontinuous ($Q$ discontinuously changes) for low temperatures
and high field amplitudes at any fixed 
frequency of the driving field (see insets of Fig. 5).
These indications suggest the existence of a 
tricritical point on
the phase boundary which separates the continuous/discontinuous nature of
transitions. Very recently, Acharyya (1999) has checked 
the existence of this tricritical point by studying the distribution $P(Q)$
of the order parameter $Q$ and the temperature variation of its
fourth order cumulant ($U_L = 1.0 - {{<Q^4>} \over {3<Q^2>^2}}$ ; 
where $<Q^n> = \int Q^n P(Q) dQ$) across the phase boundary.

 Notwithstanding the fact that the response magnetisation $m(t)$ is 
not necessarily sinusoidal,
 although it is periodic with
 the same frequency as that of the external sinusoidally varying field
$h(t)$,
 Acharyya and Chakrabarti (1994,
 1995) defined an {
\it AC susceptibility} $\chi = m(t)/h(t) \sim (m_0/h_0) \exp
 (-i\phi)$, assuming a representation $m(t) = m_0 \exp [i\omega(t - \tau_
 {eff})]$ for $h(t) \sim h_0 \exp (i\omega t)$. Here, $m_0$ corresponds
 to the amplitude of the response magnetisation and $\phi \equiv \omega
 \tau_{eff}$  is
 the effective phase lag (given by the effective delay time 
 $\tau_{eff}$ of the response $m(t)$ compared to the field $h(t)$).
 Their Monte Carlo study for two and three dimensional lattices 
 on the temperature variation of this complex 
susceptibility (moduli of the
real
 part $\chi ' = (m_0/ h_0) \cos [\omega \tau_{eff}]$ and the 
 imaginary part $\chi '' = (m_0/h_0) \sin [\omega \tau_{eff}])$ of
 an Ising model in a periodically-varying  external  field shows that 
 the prominent peak in $\chi''$ (and dip in $\chi'$) occurs
 at $T_d$ as one crosses the dynamic phase  boundary ($Q$ $\neq$ 0 for
 $T < T_d(h_0,\omega)$  and $Q$ = 0 for $T \geq T_d$). 
 It may be noted that this AC susceptibility
does not (directly) measure any asymmetry of the $m-h$ loops as
$Q$ measures: it essentially measures the delay $\tau_{eff}$ in $m(t)$
compared to $h(t)$. Yet, the AC susceptibility  measurements give
precisely the same dynamic symmetry breaking transition, where they
show peaks or dips. This feature of
the AC susceptibility is of course
 observed only for small values of $h_0$
(and large values of $T$), where the dynamic transition is
continuous.

Recently, the relaxation behaviour of the dynamic order parameter has
been studied (Acharyya 1997a). Starting from any arbitrary initial 
state in the Monte Carlo study, one can study the relaxation behaviour 
of  $Q(p)$, which represent the value of  $Q$ for
the $p$-th cycle. Acharyya showed that, similar to the critical slowing
down behaviour in static critical phenomena, here also
the dynamics becomes extremely slow (in $p$)
 as one approaches the dynamic
phase boundary $T_d(h_0, \omega)$. He also showed (Acharyya 1997b)
that the 
fluctuations $\delta Q$ in $Q$, over its average value, tend to diverge
as one approaches the phase boundary (see Fig. 6). 
Very recently, Sides et al (1998c) applied a finite size scaling 
method{\footnote [3] {In normal (static) critical phenomena, as one 
approaches the transition point $T_c$, the correlation length $\xi_s$
diverges: $\xi_s \sim |T - T_c|^{-\nu_s}$. In a finite system of 
linear size $L$, one observes a pseudo-critical point at $T_c^e$
where $\xi_s$ becomes equal to $L$ : 
$T_c^e - T_c \sim L^{-1/\nu_s}$. The finite size scaling method
utilises this scaling property of
$T_c^e$ with $L$ and 
estimates the values of the correlation length and other exponents.}}
to the Monte Carlo results  for the dynamic 
transition in the two dimensional Ising model and
estimated the values of several exponents. In particular, their
findings suggest that the correlation length $\xi \sim |T - T_d|^
{-\nu}$ for this dynamic transition grows
 with an exponent value $\nu \simeq $ 1.1, slightly higher than  
the corresponding static transition exponent value ($\nu_s$
= 1) in two dimensions.

\subsection {Due to pulsed and stochastic
fields}

\noindent Since the oscillating field can be decomposed 
structurally into
successive applications of {\it positive} 
and {\it negative} field pulses, 
considerable insights can be
gained by  investigating  the nature of the  dynamic response to
isolated pulsed fields. For the individual 
application of pulses, the
``positivity'' or ``negativity'' corresponds to 
fields favouring or competing with
the existing order before the application of the pulse respectively. 
Another interesting 
study of dynamic response has been in the case 
of a stochastically varying
(in time) field, where the entire extended 
system is in a (spatially) uniform 
field which varies randomly in time. 
The relaxation time of the
extended system being much larger than 
the time unit over which the
field changes randomly, a dynamically broken 
symmetric phase again
appears.

\bigskip

\noindent {\it (a) Response to a positive pulse field:}

\noindent Acharyya et al (1997) 
studied the response of the
kinetic Ising model (1a) to a pulsed field
$h(t) = h_p$ for $t_0 \le t \le t+\delta t$ and $h(t) = 0$ elsewhere.
Here $h_p$ is in the direction of 
the existing magnetic order $m_e (T)$
for $T < T_c$ and $t_0$ is much larger than the relaxation time of
the unperturbed system at that temperature. 
The behaviour of the response
magnetisation $m(t)$ was studied as a function 
of temperature $T$, pulse height 
$h_p$ and pulse
width $\delta t$, using Monte Carlo simulations and also 
solving numerically the mean field equation 
of motion (2), with $h(t)$ as
given above. In particular, the response contribution 
to the magnetisation was 
characterised by a (maximum) height $m_p$  and a width $\Delta t$.
For
small values of $h_p$, the width ratio $R \equiv \Delta t/\delta t$ was
found to diverge at the critical point $T_c$ of the unperturbed system,
while the pulse susceptibility $\chi_p \equiv m_p/h_p $ was found to show
a finite peak at an {\it effective}
 transition temperature $T_c^e \ne T_c$. It
was found that the peak height increases and the peak position $T_c^e$ 
approaches  $T_c$ as the pulse width $\delta t$ increases: $T_c^e =
T_c + {\rm C} (\delta t)^{-x}; ~ x \simeq 0.5$ in two dimensions 
(Acharyya et al 1997). It was argued that this effective transition 
with a finite peak at $T_c^e$  occurs due to a {\it finite time} effect,
similar to the finite size effect in
static  critical phenomena (Fisher 1964;
see footnote 3) : one
observes an effective critical behaviour here when the width of the time
window $\delta t$ becomes equal to the relaxation time of the order of
$\xi_s^z \sim |T_c^e - T_c|^{-\nu_s z}$, or $T_c^e \sim T_c + {\rm C}
(\delta t)^{-x}$ with $x = 1/\nu_s z$, where $\nu_s$ and $z$ denote the
standard correlation length  and  dynamic exponents
respectively for the  equilibrium
transition  at $T_c$. Indeed, the value of $\nu_s z $ is about
2.0 for 
two-dimensional Ising systems.

Study of the response to such pulsed 
perturbations in various self-organised
critical models can help to predict 
the critical point  before the
critical threshold is reached
(Acharyya and Chakrabarti 1996). 
Many of the catastrophic disasters occur dynamically
in a self-organised way and the disaster point (threshold
stress or time) may often be 
identified as a  critical point. Naturally, the  estimate
(prior to the disaster)  of such
self-organised critical points are extremely useful.
Since most  such dynamically self-organised  models respond to the
appropriate perturbations, which in turn 
accumulate to take the system towards
the critical point, the perturbations to test the interval from the 
criticality must necessarily be local in space and time
(i.e., pulsed). Measuring the
characteristics of the response to 
local pulsed perturbations, one can
extrapolate or predict the imminent disaster point (time) in e.g., some
self-organised criticality models of earthquakes (see e.g., Chakrabarti
and Benguigui 1997).

\bigskip

\noindent {\it (b) Response to a negative pulse field and  the
magnetisation-reversal transition:}

\noindent Misra and Chakrabarti (1997, 1998) studied
 the response of the
kinetic Ising model (1a) to a negative 
pulsed field $h(t) = - h_p$ for
$t_0 \le t \le t_0 + \delta t$ and $h(t) = 0$ elsewhere. 
They  used
Monte Carlo simulations and  numerical solutions of the 
mean field equation of
motion (2) with the above form of $h(t)$. 
Due to the application of the
negative field pulse $h_p$, which  opposes 
the existing order characterised
by the equilibrium magnetisation $m_e(T)$ for $T < T_c$, 
the down-spin
domains start growing and continue 
until the field is withdrawn. 
Depending on the average
magnetisation $m(t_0 +\delta t)$ at the time of the field withdrawal, 
the domains
either grow further (for
negative values of $m(t_0 +\delta t) $) to reach eventually  the
other equivalent ordered phase with reversed 
magnetisation $- m_e$ or they 
settle down (for 
positive values of $m(t_0+\delta t) $) to the 
original ordered phase with
magnetisation $m_e$. This transition between the two well-known 
equilibrium states, driven by pulsed fields 
competing with the existing
order of any of the equivalent states, 
is essentially dynamic in nature
and has  interesting properties.

Misra and Chakrabarti studied the phase diagram in the 
$h_p - \delta t$ plane for this dynamic 
magnetisation-reversal transition (from the
phase with equilibrium magnetisation $m_e(T)$ 
before the application of the
negative pulse, to the equivalent phase 
with magnetisation $-m_e(T)$
after the application of the field) 
for different temperatures
$T$ below $T_c$. The phase boundary here gives 
the minimal combination
of the negative pulse strength ($h_p$) and 
width ($\delta t$) required
for the magnetisation-reversal transition. 
 It is seen 
that the average magnetisation $m(t_0 + \delta t)$ at the time
of withdrawal of the pulse is the appropriate order-parameter for
such a transition and the average 
relaxation time required for $m(t)$ to
settle to the final equilibrium magnetisation $m_e(T)$ or $-m_e(T)$
diverges as one approaches the phase boundary 
(from either side, at any
fixed temperature below the static critical temperature $T_c$). This
observation  strongly indicates  the thermodynamic nature of this
dynamic transition under a negative pulsed field.

We discussed above such  magnetisation-reversal 
transitions in the presence
of fluctuations (e.g., in the Monte Carlo studies). One can also study
the mean field version of such a transition, where some analytic 
estimate can be made for the phase boundary. It may be noted that
a major qualitative difference exists between the phase diagrams of the
respective cases: while in 
the presence of fluctuations $h_p (T, \delta t)
\rightarrow 0$ for the phase boundary for the pulse width $\delta t 
\rightarrow \infty$, $h_p(t, 
\delta t) = h_c(T)$ for $\delta t \rightarrow
\infty$ in the mean field case where the static coercive field $h_c(T)$
is nonvanishing for $T < T_c$. For small $\delta t$, the mean field
phase boundary can be estimated approximately by solving the linearised
mean field equation (4) with $h(t) = - h_p$ for $t_0 \le t \le t_0 +
\delta t$  and $h_p = 0$ elsewhere. This linearisation is most 
appropriate for $t \lesssim t_0 + \delta t$, and the solution can be
written as
$$ m(t) \sim \left(m_e +{h_p \over {\epsilon T}}\right) 
\exp [-\epsilon (t -t_0)]
- {h_p \over {\epsilon T}}; ~~~\epsilon = - {1 - T \over T}. $$

\noindent One can, in fact, check this 
solution by direct substitution in
(4), with the above form of $h(t)$. The magnetisation-reversal
 transition occurs if
$m(t_0 + \delta t) \le 0.$ The phase boundary 
can therefore be obtained from
the solution of $m(t_0 + \delta t) = 0$. This gives
$h_p \delta t \simeq m_eT$, for the equation of the 
phase boundary for $T \simeq T_c = 1$ and small $\delta t$
(or large $h_p$). In this region,
the above phase boundary equation in fact agrees fairly well with  
even the Monte Carlo phase diagram. When
the contributions of fluctuations become important, the above mean 
field theory fails. If $h_p \rightarrow 0$ (as in
the large $\delta t$ region of the phase diagram), one can again use the
picture of nucleation of a single domain.  
Equating the  growth rate given by the 
Becker and D\"oring nucleation rate $I$ (discussed in section III) with
the inverse pulse width, one gets $\delta t \simeq \exp (1/h_p^{d-1})$,
suggesting $h_p$ln$\delta t =$ constant along the phase boundary
in two dimensions. This  agrees fairly well with the
Monte Carlo estimated phase diagrams in the same dimensions, 
except at very low temperatures where larger fields induce
crossover to multi-domains
(Misra and Chakrabarti 1997, 1998, Acharyya and Stauffer 1998).

\bigskip

\noindent {\it (c) Dynamic transition due to random 
(time varying) fields:}

\noindent Very recently, an interesting version of this dynamic phase
transition has been predicted for a highly
anisotropic (Ising-like) magnetic system
when the external field on the system varies in time stochastically.
 Acharyya (1998a) studied the long time
response (magnetisation) of a kinetic Ising system represented by the
Hamiltonian (1a) when the uniform field over the sample $h(t)$ 
varies randomly in time with a white distribution bounded between
$+h_0/2$ and $-h_0/2$. In a Monte Carlo simulation study in two
dimension, Acharyya  studied the nature of the response magnetisation
(see Fig. 7a,b) and determined 
the dynamic order parameter $Q (= (1/\tau)\int 
_0^{\tau} m(t') dt'; ~ \tau
>> 1)$ given by the long-time average (over the active duration
of the magnetic field) of magnetisation.
It was found that $Q$ assumes nonzero values below a phase boundary
line in the $h_0 - T$ plane, and vanishes continuously at the
transition boundary (see Fig. 7c). Again, the dynamic symmetry
breaking transition occurs due to the competing time scales; the
relaxation time of the many-body system being larger than the switching
time of the random field. Such a dynamic 
transition is again a nonequilibrium
transition, very similar to that for oscillating fields discussed
  in Section IV(A).  It may be mentioned
that, in  a slightly different context, a discrete map
version of the mean field equation of motion (2) with similar
stochastically varying field $h(t)$ was analysed recently by
Hausmann and Ruj$\acute {\rm a}$n (1997).

\bigskip

\section {Concluding remarks}

\noindent The equilibrium response of cooperative systems to 
external perturbations is  now well understood. The nonequilibrium
(yet steady state) responses of such systems to time driven 
perturbations are extremely important technologically and involve
intriguing physics. In  cases of magnets under oscillating
fields, although some of the phenomena have long been used in
technology, little was understood until very recently. The scaling
properties of dynamic hysteresis and the spontaneous symmetry
breaking dynamic transition are now somewhat  
established and partly understood. The role of stochastic resonance
in such dynamic hysteresis phenomena are, although identified, not
fully investigated or understood.

The nature of the dynamic transition 
due to negative pulses, or that
occurring for stochastically varying fields 
in time, are not yet adequately investigated.
Although the 
existence of these transitions are fairly well established, 
their detailed nature remains poorly understood so far. 

Although it is a bit surprising  that the interest
in such commonly encountered technologically important 
problems came very late
in the day, the subject has developed momentum in the
last few years and a lot of interesting 
physics has already been developed. 
We conclude with the
hope that this brief introductory review  on the exciting developments
on these rather simple and closely encountered dynamical problems
in extended systems
will inspire  further investigations. No doubt, soon
one can expect  a maturity
of the field, leading to
new physics and to better utilisation of its technological potential. 

\bigskip

\noindent {\bf Acknowledgements:}
We thankfully 
acknowledge the collaborations with J. K. Bhattacharjee, A. Misra,
A. K. Sen, D. Stauffer  and R. B. Stinchcombe.
We are thankful to C. Dasgupta, D. Dhar, R. Pandit, M. Rao,
P. A. Rikvold, D. Stauffer and
G C Wang 
for  useful comments and suggestions. We would like to thank
C. Denniston, A. Dutta, A. Misra,
 S. Sil and R. B. Stinchcombe for their comments on  the manuscript. We are 
grateful to U. Nowak and M. Staats for their help in using
the figure-scanner. Financial supports from the SFB 341 (Cologne
University),
and Graduiertenkolleg, Struktur und Dynamik Heterogener
Systeme (Duisburg University), are 
 gratefully acknowledged by MA. BKC is grateful to the 
Royal Society, London, for supporting his visit to the Physics
Department, Oxford, where a part of this work was done
and the final version of this paper was written.

\bigskip

\centerline {\bf Figure captions}

\bigskip

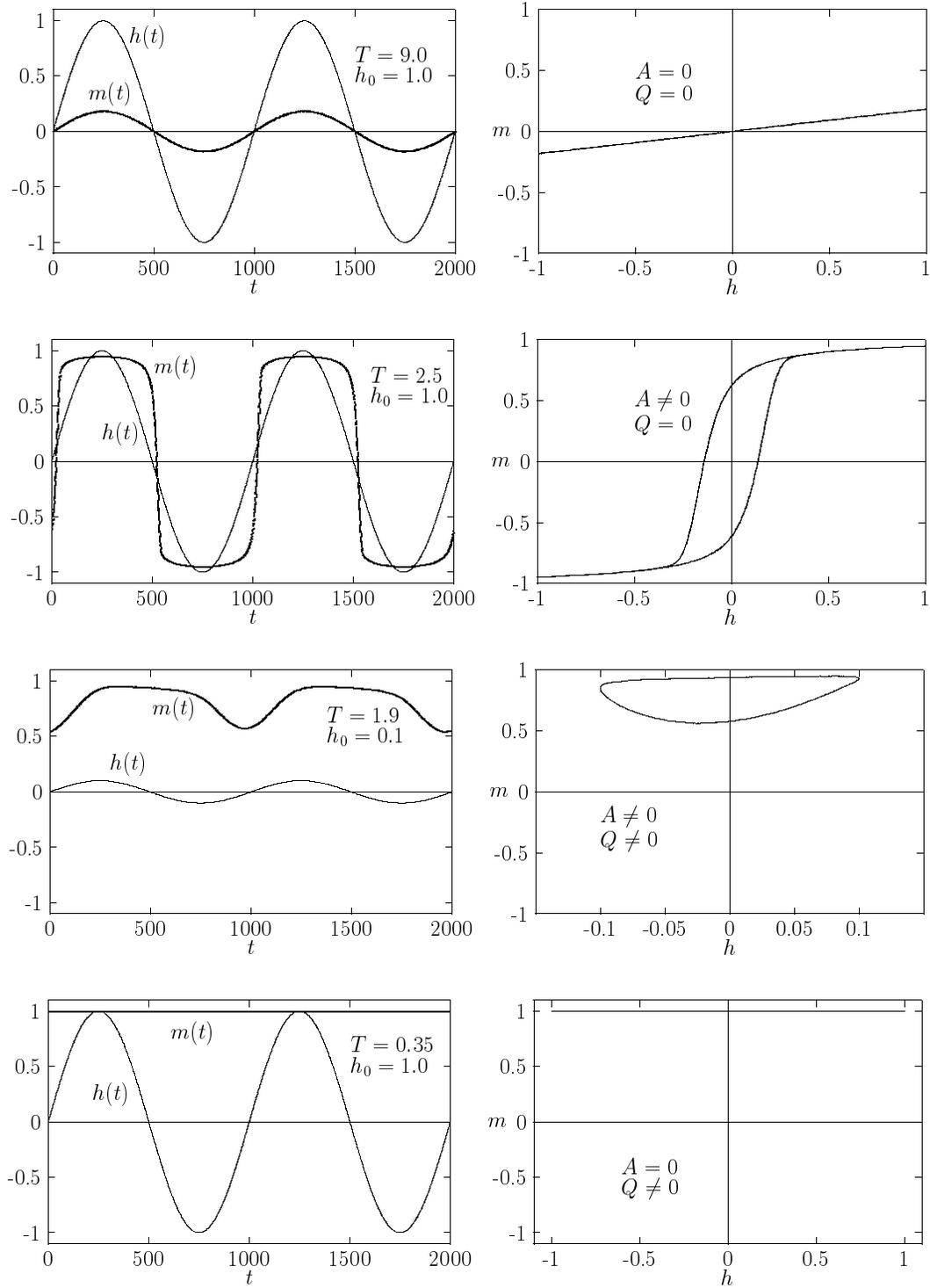
\noindent {\bf Fig. 1.} Schematic time variation
of the response magnetisation $m(t)$ compared to that of 
the oscillating field $h(t)$ for different values of frequency
$\omega$ and amplitude $h_0$
of the oscillating field and temperature $T$ of the system. 
The results are in fact actual Monte Carlo simulation
results for an Ising model on a square lattice 
 with the values for $h_0$ and $T$ as indicated in
the Figures. The Figures on the right hand side  show the 
corresponding $m-h$ loops. The values for loop area $A$ and the dynamic
order parameter $Q$ are also indicated in these figures. As one
can see, the first two cases correspond to $Q =$ 0, while the
other two correspond to dynamically broken symmetric phase (with
$Q \ne$ 0). The first figure and the last correspond to $A \simeq$
0, while the middle two correspond to nonvanishing $A$.

\bigskip

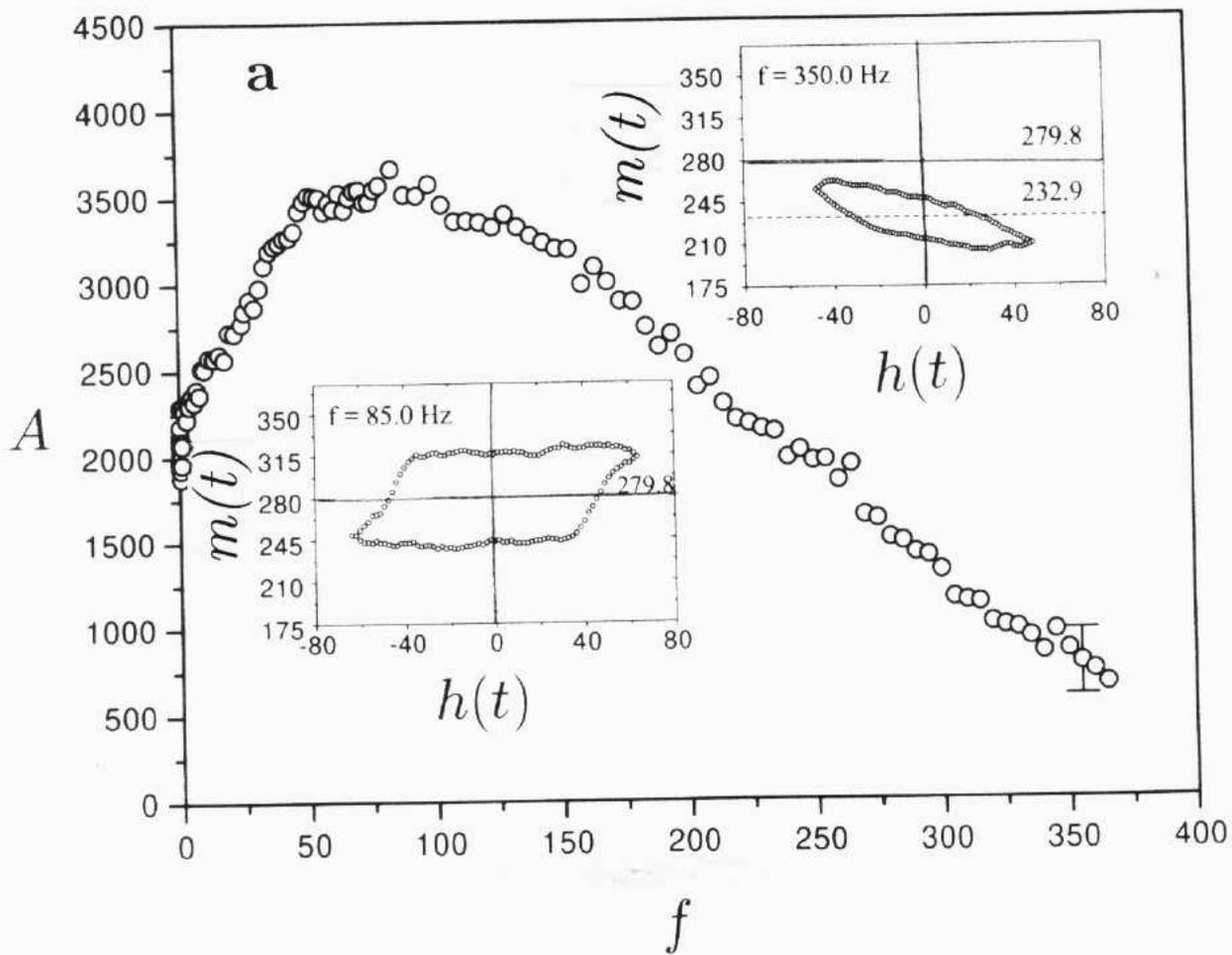
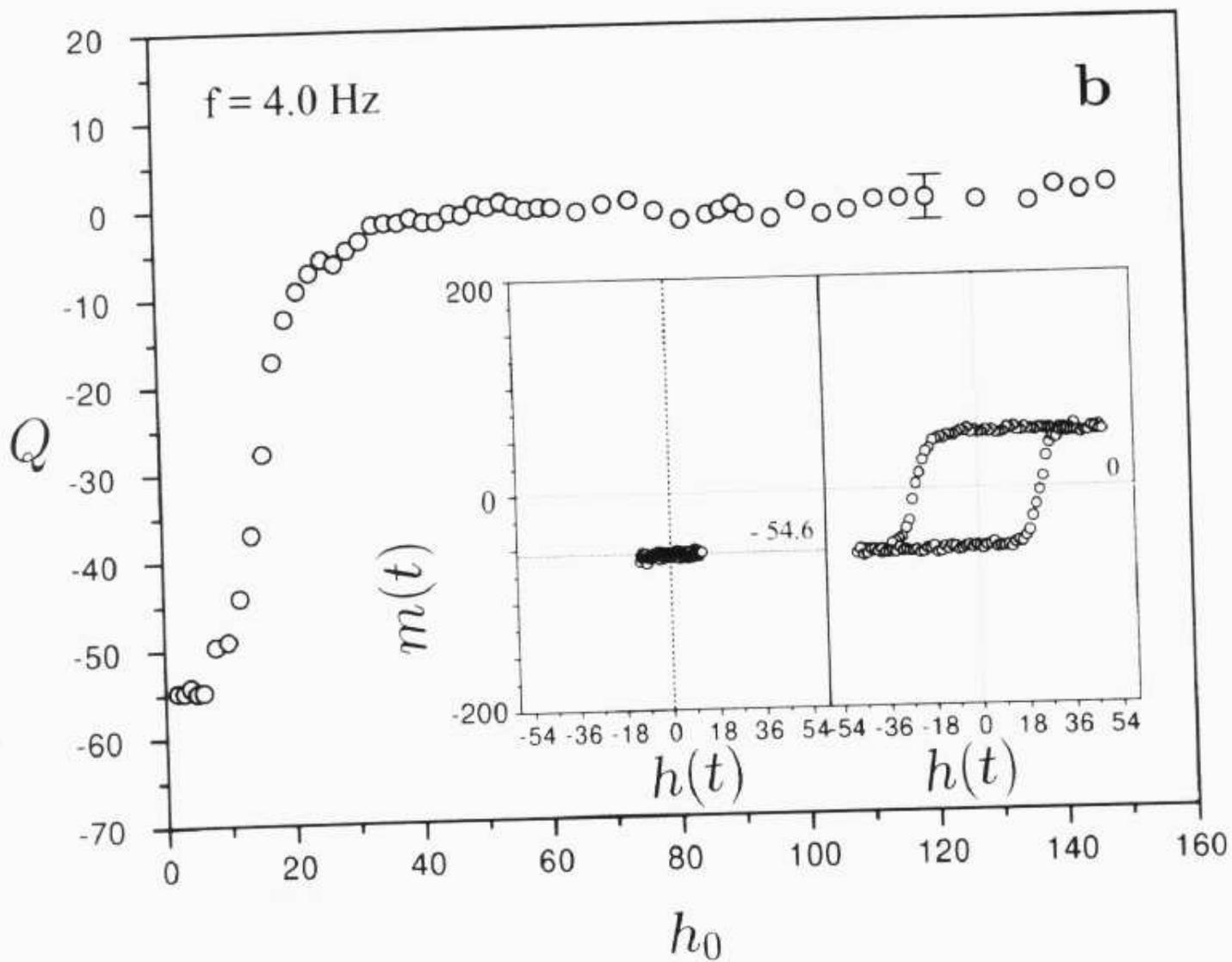
\noindent {\bf Fig. 2.} Experimental results for the dynamic hysteresis
loop area $A$ and the dynamic order parameter $Q$ (Jiang et al. 1995).
 {\bf (a)} The  results for the
loop area $A$ as a function of frequency $f$
is plotted at a fixed ac current of 0.4 
Amp. The direction of the magnetic field
is  parallel to the film plane. 
The insets show plots of $m-h$ loops for the following particular values of
the field amplitudes $h_0$: (i) $h_0 = 48.0$ Oe (top inset) and (ii) $h_0 = 63.0$ Oe (bottom inset).
{\bf (b)} The dynamic order parameter $Q$, i.e,
the average magnetisation over a cycle, is plotted 
against the field amplitude at a fixed frequency $f$ = 4 Hz.
The insets show plots of $m-h$ loops for the following particular values of
the field amplitudes $h_0$: (i) $h_0 = 48.1$ Oe (right inset) and (ii) $h_0 = 12.0$ Oe (left inset).

\bigskip

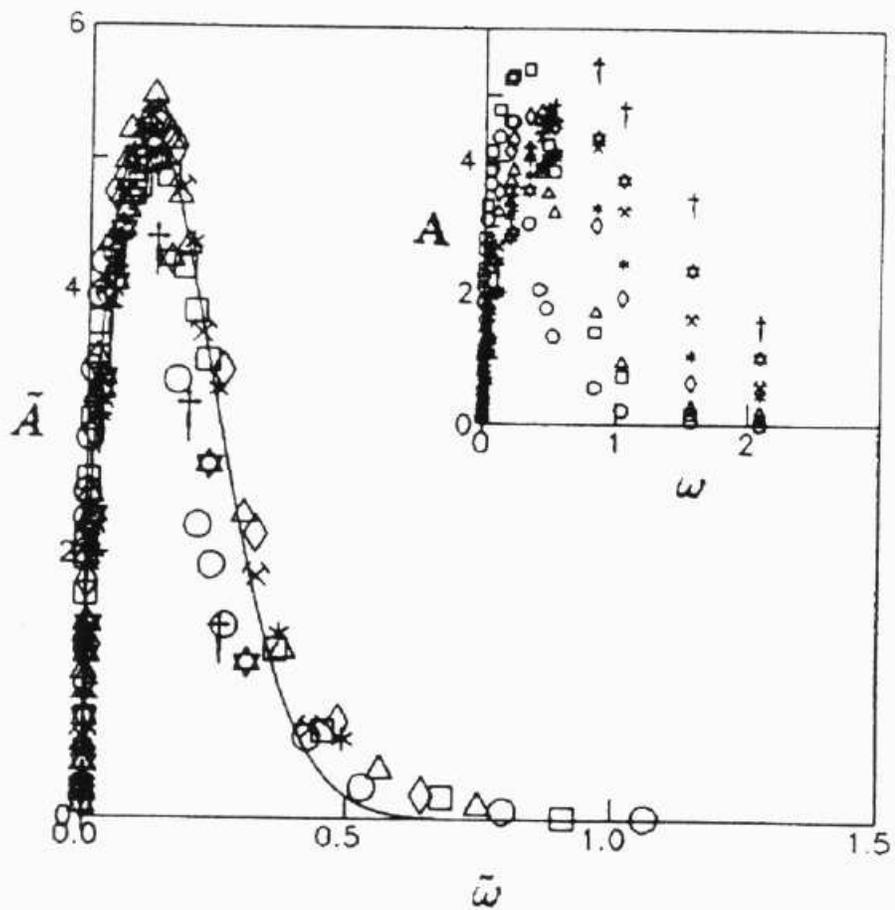
\noindent {\bf Fig. 3.} Variation of scaled loop area $\tilde 
A =  Ah_0^{-\alpha}T^
{\rho}$    with    the    scaled    frequency    $\tilde    \omega    =
\omega/h_0^{\gamma}T^{\delta}$
for the Monte Carlo data in $d = 2$. The inset shows the variation of $A$
with $\omega$ at different $h_0$ and $T$.
Different
symbols  correspond  to  different  $T$ and $h_0$: ($\circ$)
$T$ = 1.5, $h_0$ =
1.25; ($\Box$) $T$ = 1.5,  $h_0$ = 1.5; 
($\triangle$) $T$ = 2.0,  $h_0$ = 1.25; ($\diamond$) $T$ =
2.0,  $h_0$  = 1.5; ($\star$) $T$ = 2.5,  $h_0$ = 1.5; 
($\times$) $T$ = 2.5,  $h_0$ =
1.75; ($\bigtriangleup\mkern-16.0mu\bigtriangledown$ )
 $T$ = 3.0,  $h_0$ = 2.0;  and 
($\dagger$) $T$ = 3.0,  $h_0$ = 2.5.  The
solid  curve  indicates  the  proposed   scaling   function  $g(
\tilde \omega)  \sim
\tilde \omega^{\beta}exp(-\tilde \omega^2/\sigma)$  with  $\beta$  = 0.3. 

\bigskip

\noindent {\bf Fig. 4.} Schematic diagram of the dynamic phase boundary in the
field amplitude ($h_0$) and temperature ($T$) plane
at a fixed nonzero frequency. The dotted line is
the boundary of the discontinuous transition and the solid line represents
the boundary of continuous transition. The small circle represents the
tricritical point (TCP). Insets demonstrate the breaking of the symmetry
of the dynamic hysteresis ($m-h$) loop due to dynamic transition.

\bigskip

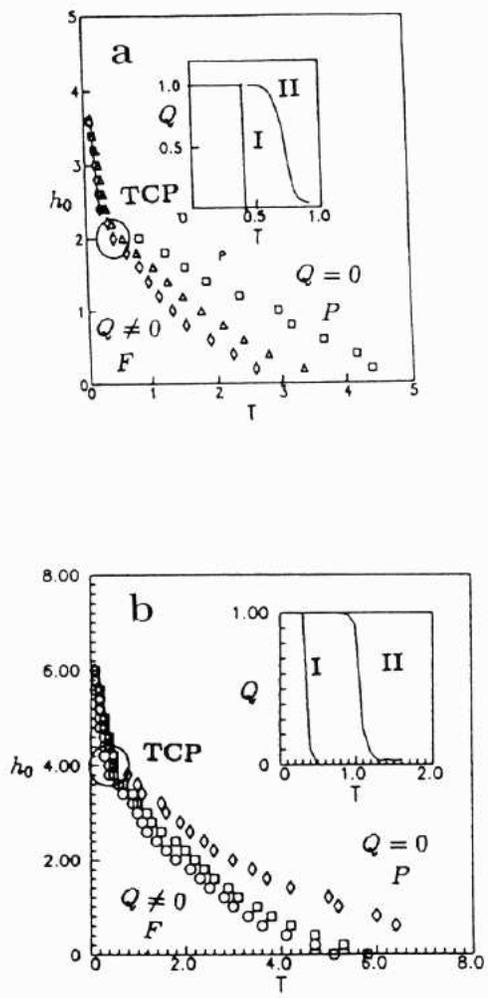
\noindent  {\bf Fig. 5.} Phase diagrams in the $h_0$-$T$ plane for various
values of $\omega$ gives the functional form of the transition 
temperature  $T_d(h_0,\omega)$  for  the  dynamic  
phase
transition: Monte Carlo results {\bf (a)} for system sizes $L = 100$ in $d = 2$, and 
{\bf (b)} for $L$ = 20 in $d$ = 3. 
Below $T_d(h_0,\omega)$, $Q$ acquires a nonzero value in
F phase  and $Q$ = 0 in P  phase.  Different  symbols  denote  different
phase  boundary  lines  corresponding  to  
different  frequencies ($\omega$):
($\Box$) $\omega$ =
0.418, ($\triangle$) $\omega$ = 0.208,
($\diamond$) $\omega$ = 0.104 in {\bf (a)}; and ($\diamond$) $\omega$
= 0.418, ($\Box$) $\omega $ = 0.202,
($\circ$) $\omega$ = 0.104 in {\bf (b)}. The locations of
the tricritical points (TCP) are indicated by the circle. 
The insets show the
nature of the  transition  just  above (I: $h_0$ = 2.2 and 4.4 in
{\bf (a)}  and {\bf (b)} respectively) and below (II: 
$h_0$ = 1.8 and 3.6 in {\bf (a)} and {\bf (b)} respectively)
the  tricritical  points
along the phase boundaries.

\bigskip

\noindent {\bf Fig. 6} Comparison
of the temperature variations of the Monte Carlo
results in $d$ = 2, 
 with $L = 100$,
 $h_0$ = 0.2 and $\omega$ = 0.063:
 $Q$ (solid line), $-dQ$/$dh_0$
($\bullet$) and
$\delta Q^2$ ($\blacktriangle$).

\bigskip
\noindent {\bf Fig. 7.} Dynamic transition due to randomly
varing fields in time.
{\bf (a, b)} Typical time variation of magnetisation $m(t)$ 
compared to that of the stochastically varying field $h(t)$
in a Monte Carlo study in $d = 2$, 
with
$ L = 100, T = 1.7$:  $h_0 = $ 3.0 for 
{\bf (a)}  and $h_0 =$ 1.0 for {\bf (b)}.
{\bf (c)} The  corresponding dynamic transition phase boundary
(separating the regions with average magnetisations $Q$  zero from
nonzero) 
in the field width ($h_0$) - temperature ($T$) plane. The 
data points are obtained using both sequential updating ($\diamond$)
and random updating ($\bullet$) in the Monte Carlo simulation.

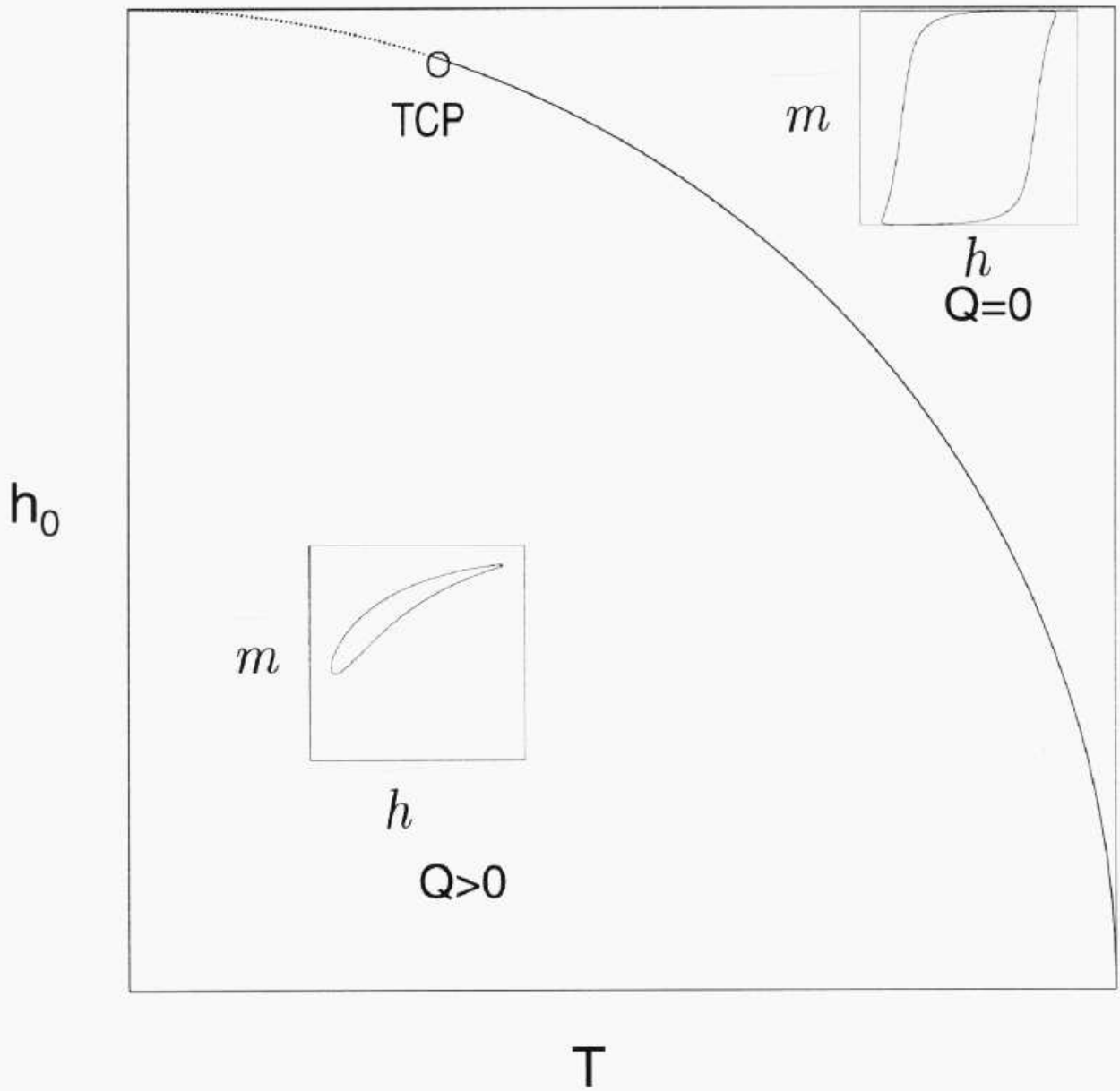
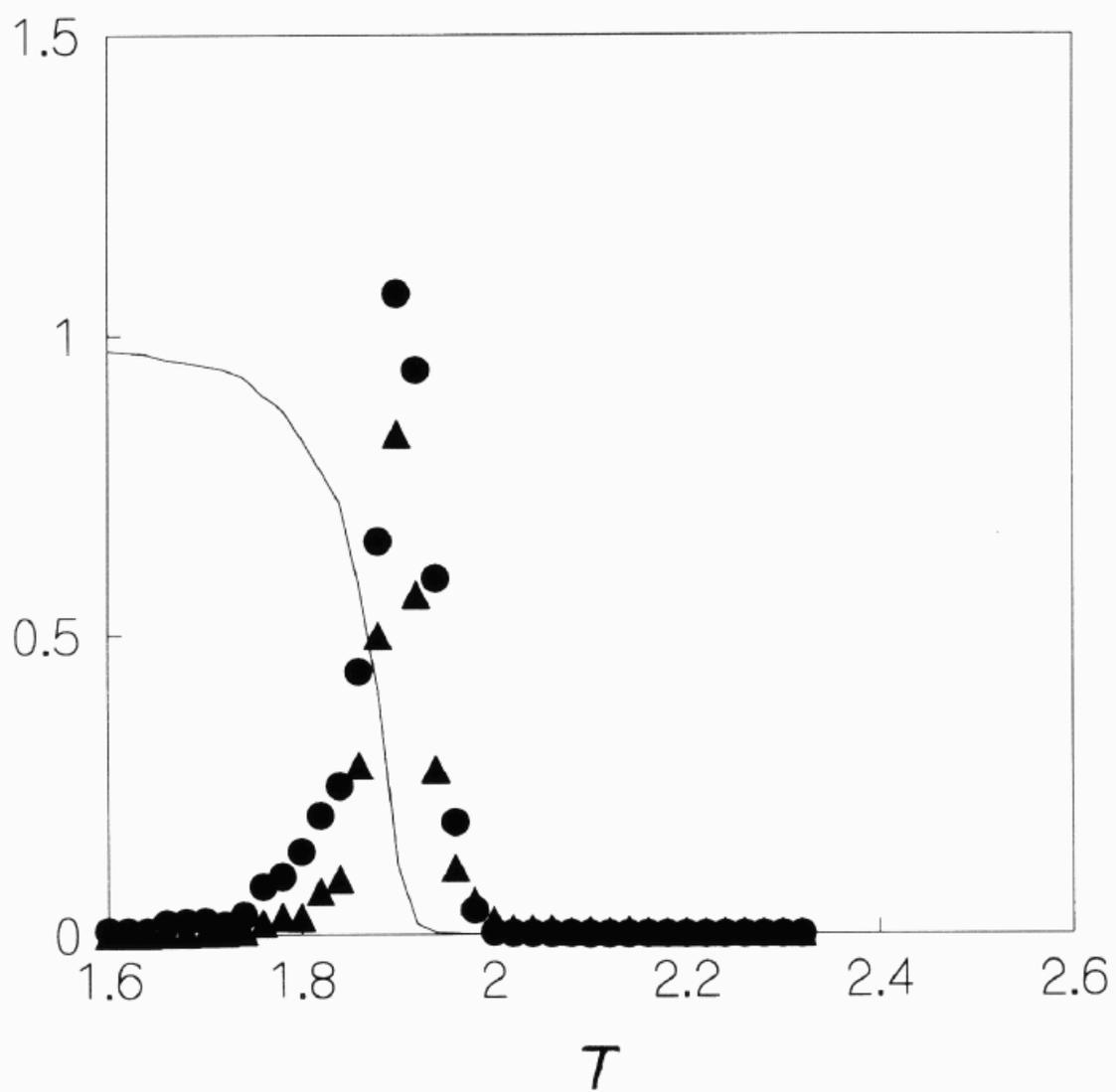
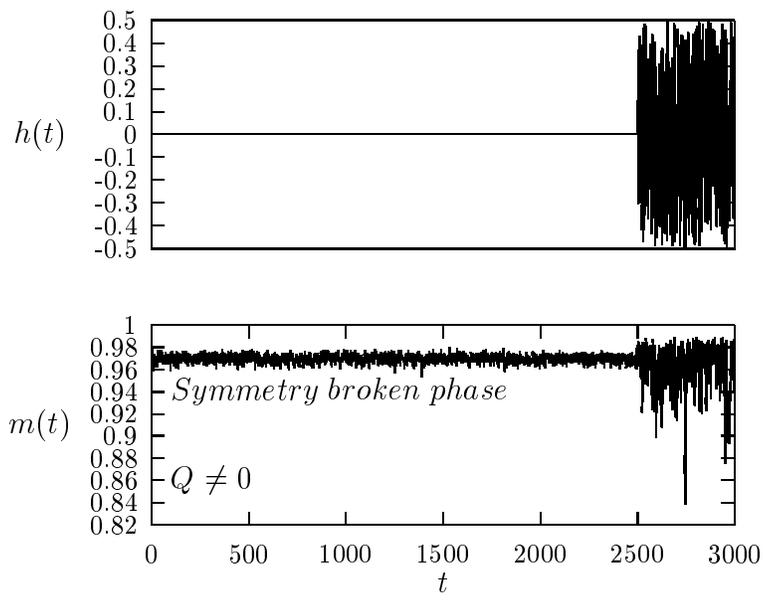
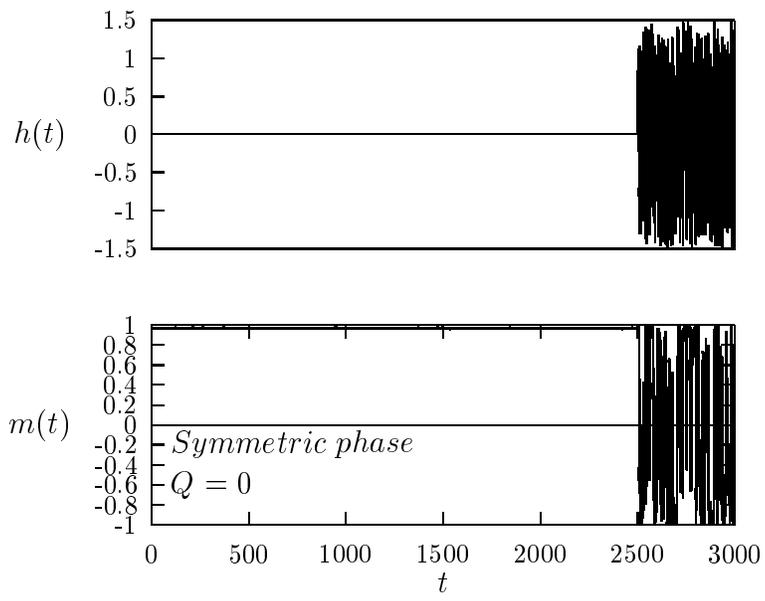
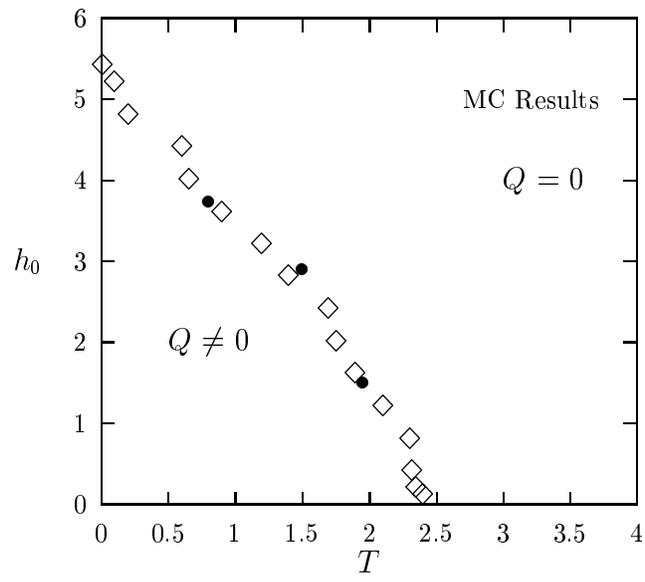
\end{document}